# Large thermoelectric figure of merit in graphene layered devices at low temperature

Daniel Olaya[1], Mikel Hurtado-Morales[1,2], Daniel Gómez[1], Octavio Alejandro Castañeda-Uribe[3], Zhen-Yu. Juang[4,5], Yenny Hernández[1*]

[1]Nanomaterials Laboratory, Department of Physics, Universidad de los Andes, Bogotá 111711, Colombia

[2]Deparment of Electronic Engineering, Universidad Central, Calle 21 # 4 – 40, Bogotá – Colombia.

[3]Department of Biomedical Engineering, Universidad Manuela Beltrán, Avenida circunvalar 60 – 00, Bogotá – Colombia.

[4]Department of Electrophysics, National Chiao Tung University, Hsinchu 30010, Taiwan

[5]SulfurScience Technology Co., Ltd., Taipei 10696, Taiwan.

## Abstract

Nanostructured materials have emerged as an alternative to enhance the figure of merit ($ZT$) of thermoelectric (TE) devices. Graphene exhibits a high electrical conductivity (in-plane) that is necessary for a high $ZT$; however, this effect is countered by its impressive thermal conductivity. In this work TE layered devices composed of electrochemically exfoliated graphene (EEG) and a phonon blocking material such as poly (3,4-ethylenedioxythiophene) polystyrene sulfonate (PEDOT:PSS), polyaniline (PANI) and gold nanoparticles (AuNPs) at the interface were prepared. The figure of merit, $ZT$, of each device was measured in the cross-plane direction using the Transient Harman Method (THM) and complemented with AFM-based measurements. The results show remarkable high $ZT$ values (0.81 < $ZT$ < 2.45) that are directly related with the topography, surface potential, capacitance gradient and resistance of the devices at the nanoscale.

Introduction

TE materials have attracted the attention of the automotive, aerospace, medical and electronic industries due to their ability to transform waste heat into electricity by the Seebeck effect[1]. The performance of these materials is determined by their figure of merit ($ZT = S^2\sigma T/\kappa$), which is a function of the Seebeck coefficient ($S$), the electrical conductivity ($\sigma$), the thermal conductivity ($\kappa$) and the temperature ($T$). The figure of merit of conventional TE materials (highly doped semiconductors) is a balance between the electrical conductivity and the thermal conductivity (connected by the Wiedemann-Franz Law)[2]. This balance leads to $ZT$ values not greater than the unit[3], which limits the use of TE materials in power generation and energy harvesting applications. $ZT$ is further limited in the presence of small temperature gradients[2] and TE materials for applications where this is the case are yet to be developed.

An alternative for increasing $ZT$ is the use of low dimensional materials as proposed by Dresselhaus and Hicks[4,5]. This improvement is due to an enhanced Seebeck coefficient, a dimension-dependent electronic density of states and a low thermal conductivity due to phonon scattering at the interfaces[6]. These findings lead to the development of nanostructured TE materials with $ZT$ values up to 2.4 (at a working temperature of ~1000 K)[7].

Carbon materials have the broadest range of thermal conductivity values reported[8]. These values go from 0.01 W/mK for amorphous carbon up to 2500 W/mK for diamond. In the case of graphite, it exhibits a high anisotropy in its thermal conductivity ($\kappa$) in the cross-plane and in-plane directions with values at around 10 W/mK and 2000 W/mK respectively. This anisotropy is also observed in the electrical conductivity ($\sigma$) of graphite with values at $3 \times 10^2$ S/m and $2 \times 10^5$ S/m in the cross-plane and in-plane directions respectively.

Low dimensional carbon-based materials, such as graphene and carbon nanotubes (CNTs) present good electrical and thermal conductivities (in-plane), which imposes an obstacle for their use as TE materials. Multilayer stacking of two dimensional (2D) materials has been proposed as an efficient route towards the enhancement of thermoelectric properties[9,10]. In particular, solution exfoliated graphene films[11-13] display low thermal conductivity in the cross-plane direction[14] resembling the reported value for graphite[8]. Conducting polymers such as PEDOT:PSS and PANI have been used for TE studies due to their preferential electrical conductivity along the polymer chain direction, their low thermal conductivity and their measurable Seebeck response[15]. Composite materials of these polymers with graphene[16,17] and carbon nanotubes[18] (CNT) have been recently prepared and this has led to improved values of $S$ and $\sigma$, which increases directly the power factor. Additionally, theoretical calculations of gold nanopillars patterned on graphene predicted the presence of a large in-plane Seebeck coefficient for such structure[19].

In this work, an approach to enhance the cross-plane figure of merit of graphene-based TE materials is proposed. Solution processing methods were used to fabricate layered devices based on electrochemically exfoliated graphene (EEG) and interlayer conducting materials, such as PEDOT:PSS, PANI and AuNPs. The TE performance of the fabricated devices in terms of the figure of merit, Seebeck coefficient and electrical conductivity is characterized in the cross-plane direction by means of THM[20]. In addition, the devices are structurally and electrically characterized at the nanoscale by AFM. The local maps of topography, surface potential, capacitance gradient and resistance are measured to study the influence of the nanostructured materials in the overall TE behavior of the devices.

### ZT measurement of layered devices via THM

Layered materials of graphene and a conductive interlayer material such as AuNPs, PEDOT:PSS and PANI, were prepared in a configuration 3:2 following the coating procedures described in the methods section (Figure 1). The thermoelectric characterization was carried out by means of THM with current pulses of 1 ms which allowed statistical analysis (7 cycles) of the extracted data (Figure 2a). Within THM, the voltage drop is divided in two regimes: a rapid voltage drop due to the resistance, $V_R$, followed by a slow decline due to the Seebeck voltage, $V_S$, as the thermal gradient across the device dissipates (Figure 2b). The turning point between the two regimes was determined using a linear and an exponential regression for each one. The ratio between $V_S$ and $V_R$ is equivalent to the *ZT* figure[21] (*ZT* = $V_S/V_R$) and can be calculated as a function of temperature (Figure 2c).

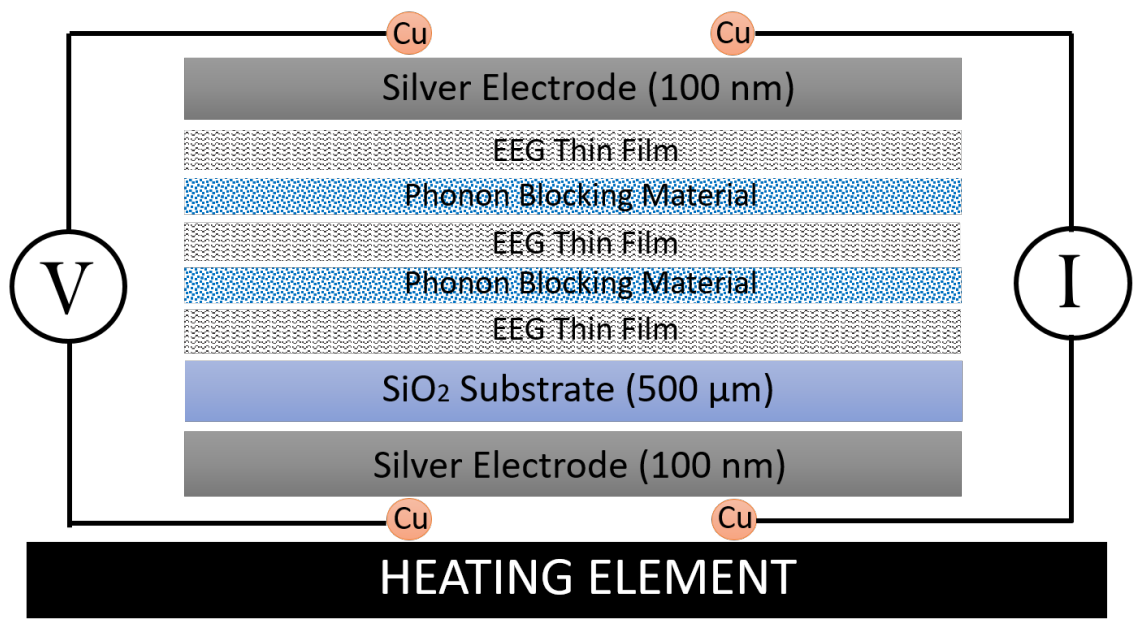

**Figure 1.** Diagram of a layered TE device (configuration 3:2).

The layered devices built with PEDOT:PSS, PANI and AuNPs have average *ZT* values of 2.45, 1.47 and 0.81 respectively (Figure 2c). The *ZT* figure of the EEG – Polymer structures is unprecedented for solution processed graphene[16,17] and when compared with $Bi_2Te_3$ based TE materials working at the same temperature range[22]. The *ZT* with AuNPs at the interface is in good agreement with the report by Juang and coworkers on CVD graphene – AuNPs heterostructures[23]. The dotted line in Figure 2c represents the ZT figure of a device composed purely by $SiO_2$ and silver contacts as a reference for the contribution of the substrate and the electrical contacts.

The semiconducting behavior of the layered devices was corroborated by plotting the maximum voltage measured as a function of temperature (Figure S1). This measurement was coupled with a comparison of the electrical resistance in the in-plane and cross-plane configurations to further corroborate the anisotropy of our devices (Figure S1).

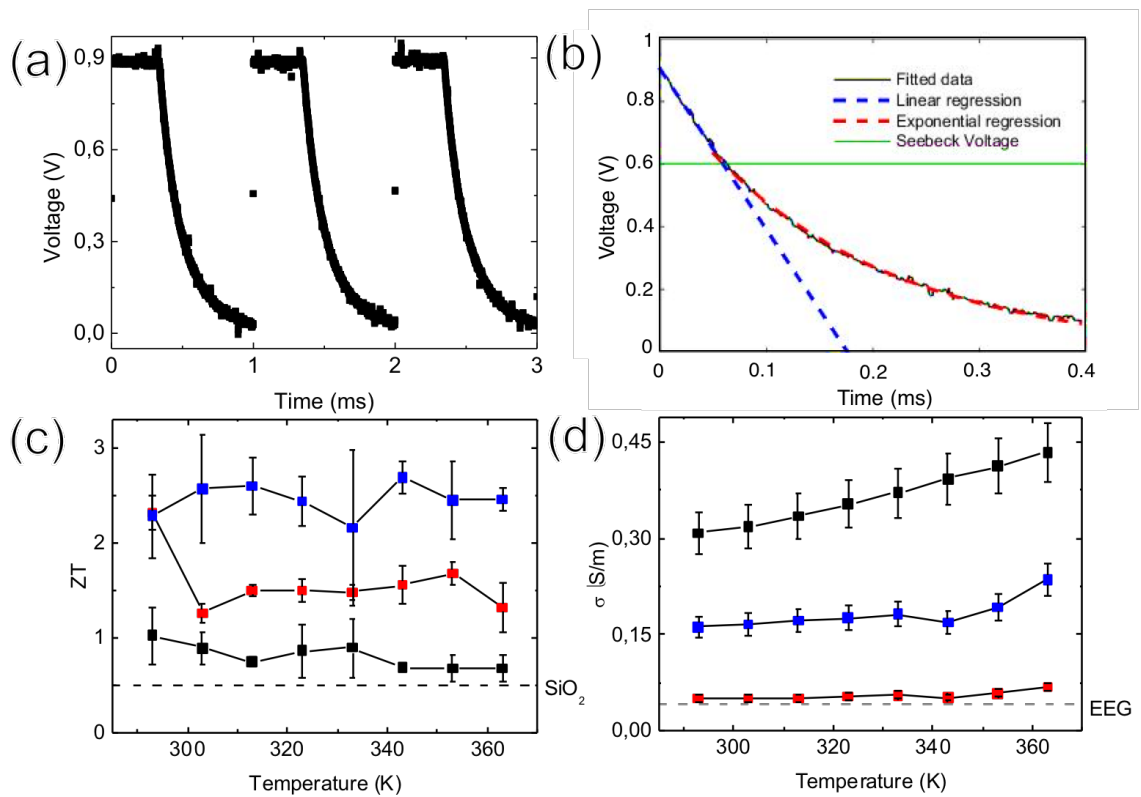

Figure 2. (a) Representative figure of a THM measurement (b) Intercept between the resistive and the Seebeck regime (c) and (d) *ZT* and σ for EEG – AuNPs (black square), EEG – PANI (red square) and EEG – PEDOT:PSS (blue square) devices as a function of temperature (configuration 3:2).

Influence of the interlayer material

One of the goals when choosing the interlayer material was to enhance the electrical conductivity across the heterostructure in comparison to a device built purely with EEG. For this reason and for their TE response[15], conducting polymers such as PEDOT:PSS and PANI were chosen. The performance of such polymers in a cross-plane configuration was measured by THM as a comparison with the EEG heterostructures. The ZT figure of these devices was ~1 which is lower when compared to the heterostructures (Figure S2). Figure 2d shows the internal conductivity of the layered devices as a function of temperature. It is important to note that even though the device with AuNPs displays higher conductivity than the one with PEDOT:PSS its *ZT* figure is what ultimately would determine the efficiency of the device. The dotted line in Figure 2d represents the average conductivity of an EEG device in comparison with the improvement of the electrical conductivity due to the presence of the interlayer materials.

The Seebeck coefficient, extracted from the THM measurement ($V_S$) and the measured temperature gradient, appear to be independent from the nature of the interlayer material (Figure 3a). It is noteworthy, however, that this response is enhanced when compared to a pure EEG device (Figure 3d). Meanwhile, the power factor ($S^2\sigma$) differs between devices at low temperature due to the difference in electrical conductivity. Even though, the AuNPs device displays a higher value than that of the devices with conducting polymers at room temperature, as the temperature increases, this difference is not as pronounced and this determines the recommended operating temperature for the TE device (Figure 3b). Although $\kappa$ was not measured directly it can be calculated by dividing the power factor and the *ZT* figure to illustrate the importance not only of having high $\sigma$ but also low $\kappa$ for the TE efficiency calculation (Figure 3b (inset)). Note that the extracted values for $\kappa$ are within the same order of magnitude to those reported for graphite in the cross-plane direction[8].

It is interesting to compare the devices with highest *ZT* and highest power factor with an EEG device. On the one hand, to explain the high *ZT* figure measured for the EEG device (Figure 3c) there are a couple things to consider: EEG has typical crystal sizes of ~5μm (Figure S3) and thin films are deposited by spray coating, up to 20nm in thickness, to assure full coverage of the substrate. The wrinkles and edges of the deposited EEG films (Figure S4) are known to contain sharp features in the electronic density of states[24] which plays an incremental role on the Seebeck coefficient[25] and in the electrical conductivity in the cross-plane direction.

On the other hand, the thermal conductivity across the EEG device is two orders of magnitude lower than the highest calculated value κ for the layered devices; hence the high *ZT* figure (Figure S4). The Seebeck coefficient of the graphene device is in the same order of magnitude of the layered devices, however its power factor is significantly lower, which could be attributed to its low electrical conductivity (Figure 3d).

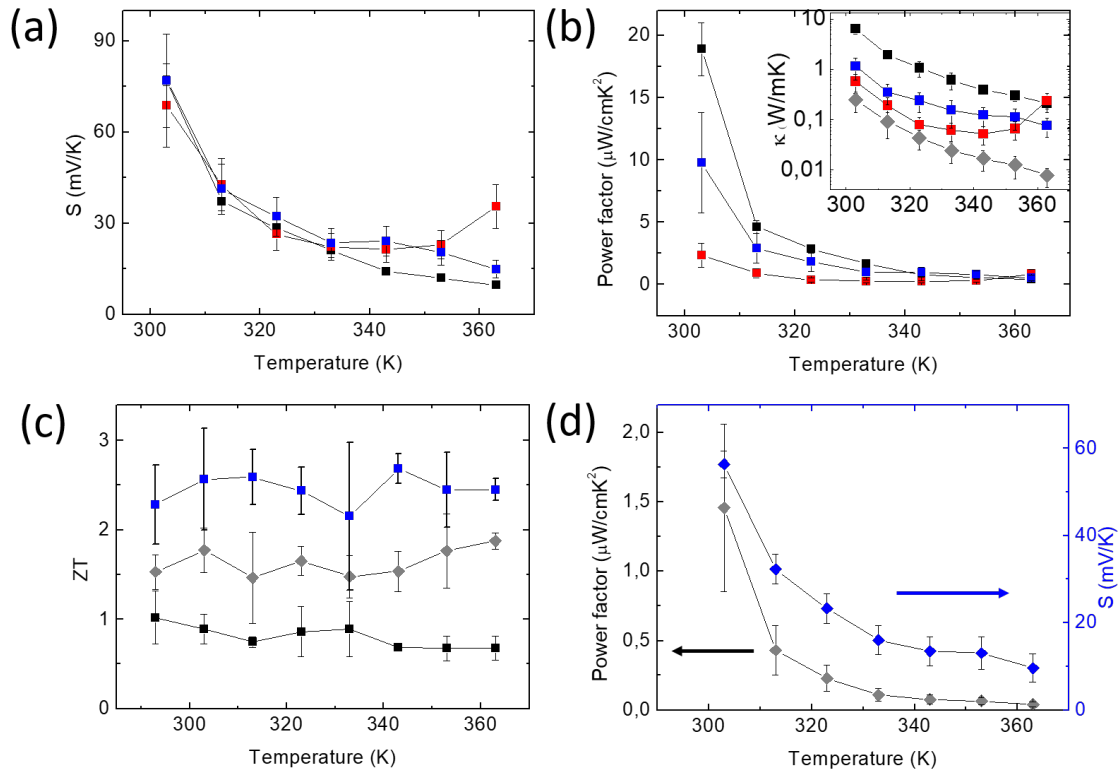

Figure 3. (a) and (b) Seebeck coefficient, power factor and κ (inset) for EEG – AuNPs (black squares), EEG – PANI (red squares), EEG – PEDOT:PSS (blue squares) and EEG (grey squares) devices as a function of temperature (c) *ZT* for EEG (grey diamonds), EEG – AuNPs (black squares) and EEG – PEDOT:PSS (blue squares) devices as a function of temperature (d) Power factor and Seebeck coefficient of EEG as a function of temperature.

### Nanoscale characterization

The nanoscale electrical characterization of the samples was divided into independent measurements of surface potential, capacitance gradient and local resistance. The surface potential images (Figure 4b) show a heterogeneous distribution of potentials, which is attributed to the difference in the work functions of the constitutive materials of each device (EEG, conducting polymer and AuNPs layers). Particularly, the EEG – PANI device exhibits extensive equipotential regions (yellow and turquoise areas) that work as potential barriers that block the electron transport in the cross-plane direction.

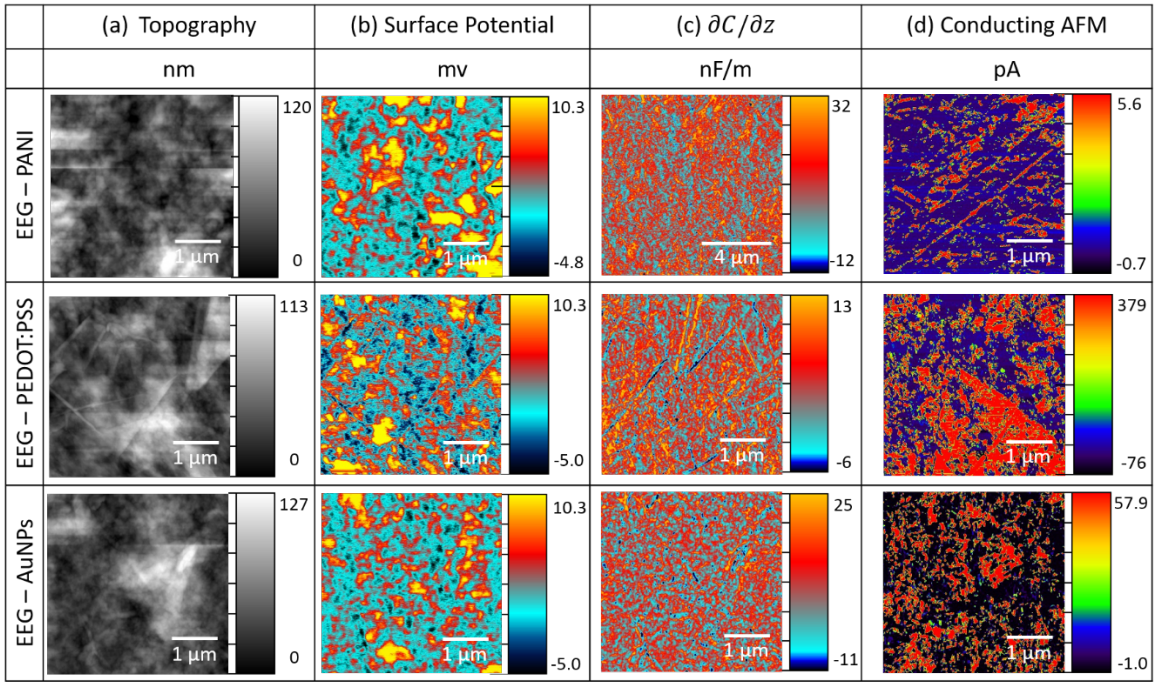

**Figure 4. (a)** Topography, **(b)** Surface potential, **(c)** $\partial C/\partial z$ **(d)** Conducting AFM of EEG – PANI, EEG – PEDOT:PSS and EEG – AuNPs.

The capacitance gradient denotes the ratio of the variations between sample capacitance and height ($\partial C/\partial z$). The high homogeneity (low contrast) of the $\partial C/\partial z$ maps in Figure 4c is a result of the uniformity in the capacitance and height of the top layer material of each device. This indicates a high quality in the multi-layering process of fabrication. Moreover, the estimation of the total equivalent capacitance based on the $\partial C/\partial z$ measurements of each device paired with impedance spectroscopy analysis[26] is a complex procedure and it is out of the scope of this paper.

To elucidate the relation of the cross-plane electronic transport with the TE parameters previously discussed, conductive AFM measurements were conducted. In the case of the EEG – PANI device the density of current paths is significantly lower than the other samples which correlates well to the electrical conductivity measurements. The EEG – PEDOT:PSS device displays a greater density of current paths when compared to the EEG – AuNPs one. This indicates more prominent Joule heating that increases $V_s$ and therefore *ZT*.

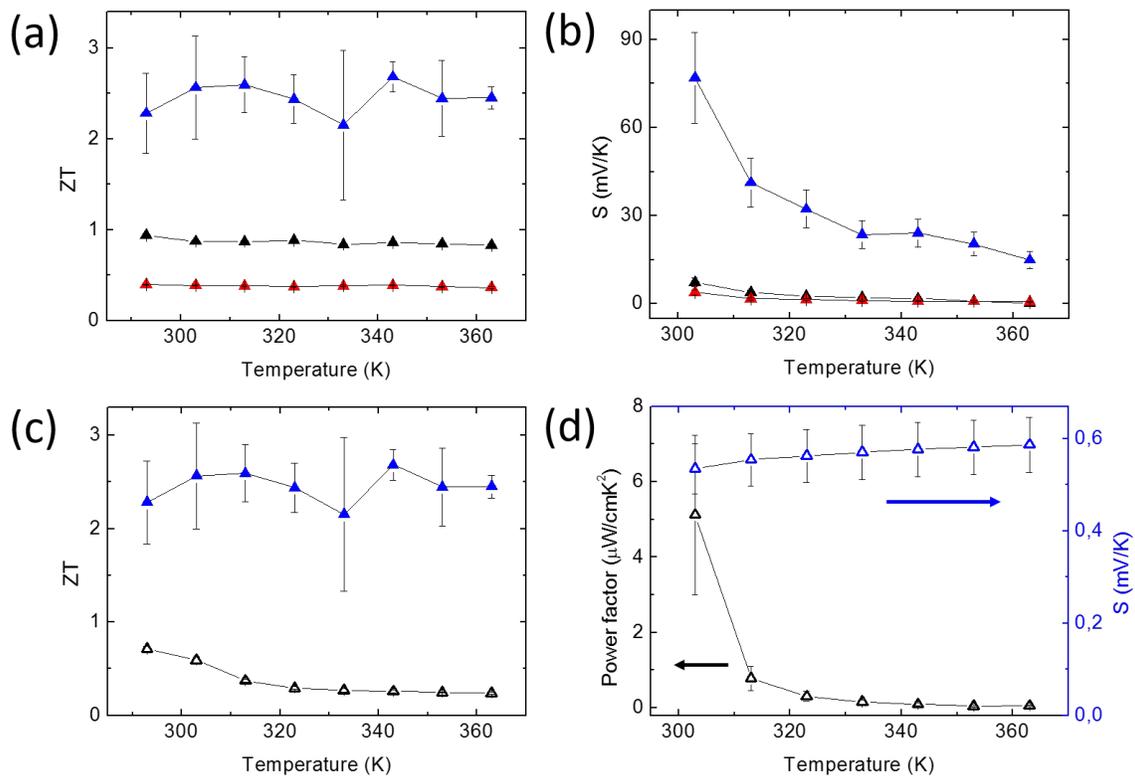

**Figure 5.** (a) *ZT* and (b) Seebeck coefficient of EEG – PEDOT:PSS (4:3) (black triangles), EEG – PEDOT:PSS (3:2) (blue triangles), EEG – PEDOT:PSS (2:1) (red triangles) as a function of temperature (c) *ZT* of EEG – PEDOT:PSS (3:2) (blue triangles) and CNT – PEDOT:PSS (3:2) (white triangles) as a function of temperature (d) Power factor and Seebeck coefficient of CNT – PEDOT:PSS (3:2) as a function of temperature.

### Layer dependence and comparison with CNT

The effect of the number of layers was also studied in configurations 2:1, 3:2 and 4:3 for the EEG - PEDOT:PSS layered device (Figure 5a). The *ZT* figure for the 2:1 and 4:3 device is lower than the 3:2 configuration and this could be associated to a lower Seebeck coefficient in both cases (Figure 5b). The value of *S* can be enhanced by increasing the number of layers (2:1 and 3:2), however this enhancement is limited by the voltage output in the Seebeck measurement (4:3). The extracted values of the thermal conductivity compared with the EEG device indicate that the inclusion of PEDOT:PSS has an asymmetric effect with the number of layers (Figure S5) which differs from previous reports on graphene only multilayer structures[27]. On the one hand, the inclusion of one layer of PEDOT:PSS (2:1 configuration) has almost no effect on $\kappa$. On the other hand, the 3:2 and 4:3 configurations displayed increased and lowered $\kappa$ values respectively. The increased value of $\kappa$ for the 3:2 configuration could be attributed to the formation of electronic percolating paths which enhances the electronic contribution to the thermal conductivity. In contrast, the lowered valued of $\kappa$ for the 4:3 configuration could be associated to phonon blocking at the interfaces. From this, it can be concluded that a close interplay between *S* and $\kappa$ would be key when designing new low dimensional TE heterostructures.

Additionally, a CNT - PEDOT:PSS device, in a configuration 3:2, was prepared with the interest of comparing it with the graphene based one. The *ZT* values measured in this case were significantly lower (Figure 5c). The low efficiency of the CNT based device could be attributed to the large anisotropy on the electrical and thermal transport of CNT which is preferential along the tube axis[8] which in this case is oriented in-plane. Additionally, a low Seebeck coefficient was measured (cross-plane) in contrast with reports on large Seebeck coeficcients in semiconducting single wall nanotube films (in plane)[28]. Further experimental efforts are needed to design carbon nanotube – graphene hierarchical structures.

Conclusion

In this work a new, low cost, method to produce high performance TE components that could be used in micro-electronic applications has been demonstrated. Graphene based devices, prepared from solution, displayed large figures of merit (0.81 < *ZT* < 2.45) when compared to their counterparts within the same temperature range (<400K). The present study correlates the thermoelectric performance of the devices with the physical properties of the interlayer material. This advancement is certainly a stepping stone towards the engineering of new advanced layered TE materials.

Methods

Graphene exfoliation, CNT dispersion and film formation

Graphene was electrochemically exfoliated from graphite foil (Alfa Aesar, 0.254 mm thick, 99.8 %) using sulfuric acid (Sigma Aldrich, 99.999%) at 0.1M at 10V. The expanded material was filtered using VVPP filters (pore size 0.1 μm), to wash residual acid on the surface. The filtered material was then dispersed in 50 ml of Millipore water via bath sonication (Branson 1800, 90 min) to finish the exfoliation procedure. Thicker graphite flakes were removed via centrifugation (Hermle Z306, 60 min at 3500 RPM). This process was conducted twice to obtain stable graphene aqueous dispersions at 0.28 mg/ml. Graphene films were formed on (1cm x 1cm) $SiO_2$ substrates (thickness 500 μm) by spray coating 1.6 ml of the dispersion at 100 °C substrate temperature.

To prepare the CNT films, an aqueous solution of sodium cholate (Sigma Aldrich) at 0.2 mg/ml was prepared. Subsequently CNT (Nano Integris, 13 – 18 nm outer diameter) were dispersed via tip sonication (QSonica, 50 W, 300 s), followed by 24h shelf decantation. This process was conducted twice to obtain stable CNT dispersions at 0.2 mg/ml. CNT films were formed on $SiO_2$ substrates by spray coating 5 ml at 100 °C substrate temperature.

AuNPs, PANI and PEDOT:PSS preparation and deposition

AuNPs (US Research Nanomaterials Inc., 14 nm) and PEDOT:PSS (Sigma Aldrich, 1.3 wt% dispersion in $H_2O$) were dispersed using a sonic tip (180 s) in Millipore water at 0.1 mg/ml and 2 mg/ml respectively. PANI (Sigma Aldrich, $M_w$ > 15.000) was dispersed in toluene using a sonic tip (300 s) at 0.2 mg/ml. AuNPs and PEDOT:PSS were deposited via spin coating (Laurell, WS-650MZ) at 3500 RPM for 40s. PANI was deposited by spray coating 1 ml at 60 °C substrate temperature .

### Production of layered devices

EEG and CNT deposition protocols were designed to attain continuous films. With the previously mentioned protocols layered devices were produced by spray coating EEG or CNT as the bottom and top layers in a typical configuration (3:2) with the interlayer materials deposited as previously mentioned. The average thicknesses was ~115 nm and ~105 for EEG and CNT devices respectively (Figure S8).

### AFM-Based Characterization

Nanoscale electrical characterization of the samples was conducted using atomic force microscopy (Asylum Research MFP3D-BIO). The surface potential, capacitance gradient ($\partial C/\partial z$) and local resistance were measured on the surface sample by means of Kelvin probe force microscopy (KPFM), second harmonic electrostatic force microscopy (2$^{nd}$ Harmonic EFM) and conductive atomic force microscopy (C-AFM) respectively. The measurements were conducted using an electrical conductive cantilever (AC240TM-R3) and applying a voltage signal (in the range of 0.02 V to 0.5V) between the tip and the sample. In addition to the electrical measurements the topography of the samples was also measured by the conventional tapping mode of AFM. All the images were taken at a scan rate of 0.5Hz and a resolution of 512 by 512 pixels.

### THM for the measurement of the figure of merit *ZT*

For THM, currents from 3 to 60 mA were injected across each device with a Keithley sourcemeter (2450) while the voltage response was sensed across them using a Tektronix (TBS1152) oscilloscope (see Figure 1). The *ZT* figure was measured at different temperatures by placing the sample on a heating element controlled by National Instruments electronics. Measurements were taken from room temperature up to 90°C ($\pm$ 0.1 °C). The electrical contacts were deposited by thermal evaporation of silver (100 nm).

THM allows the determination of the Seebeck voltage, while the temperature difference between the cold and hot side of each device is measured by thermocouples (National Instruments, k-type). These two measurements are essential to calculate the Seebeck coefficient. The internal electrical resistance is determined with the maximum voltage generated by the injected current.

## Author contributions

Y.H. and Z-Y. J. conceived the experiments. D. O., M. H-M. and D.G. fabricated the devices. D. O. and D. G performed the experiments. A. C. performed the AFM measurements. All the authors discussed the results. Y.H. and D. O. wrote the manuscript and all the authors contributed to it.